\documentstyle[11pt]{article}

\begin{document}

\newcommand{\beq}{\begin{equation}}
\newcommand{\eeq}{\end{equation}}
\newcommand{\bea}{\begin{eqnarray}}
\newcommand{\eea}{\end{eqnarray}}

\newcommand{\chii}{\raise.5ex\hbox{$\chi$}}
\newcommand{\Z}{Z \! \! \! Z}
\newcommand{\R}{I \! \! R}
\newcommand{\N}{I \! \! N}
\newcommand{\C}{I \! \! \! \! C}

\newcommand{\noi}{\noindent}
\newcommand{\vs}{\vspace{5mm}}
\newcommand{\ie}{{${ i.e.\ }$}}
\newcommand{\eg}{{${ e.g.\ }$}}
\newcommand{\ea}{{${ et~al.\ }$}}
\newcommand{\hf}{{\scriptstyle{1 \over 2}}}
\newcommand{\ih}{{\scriptstyle{i \over \hbar}}}
\newcommand{\hi}{{\scriptstyle{ \hbar \over i}}}
\newcommand{\itwoh}{{\scriptstyle{i \over {2\hbar}}}}
\newcommand{\dbrst}{\delta_{BRST}}

\newcommand{\deder}[1]{{ 
 {\stackrel{\raise.1ex\hbox{$\leftarrow$}}{\delta^r}   } 
\over {   \delta {#1}}  }}
\newcommand{\dedel}[1]{{ 
 {\stackrel{\lower.3ex \hbox{$\rightarrow$}}{\delta^l}   }
 \over {   \delta {#1}}  }}
\newcommand{\dedetwo}[2]{{    { \delta {#1}} \over {   \delta {#2}}  }}
\newcommand{\dedetre}[3]{{ \left({ \delta {#1}} \over {   \delta {#2}}  
 \right)_{\! \! ({#3})} }}

\newcommand{\papar}[1]{{ 
 {\stackrel{\raise.1ex\hbox{$\leftarrow$}}{\partial^r}   } 
\over {   \partial {#1}}  }}
\newcommand{\papal}[1]{{ 
 {\stackrel{\lower.3ex \hbox{$\rightarrow$}}{\partial^l}   }
 \over {   \partial {#1}}  }}
\newcommand{\papatwo}[2]{{   { \partial {#1}} \over {   \partial {#2}}  }}
\newcommand{\papa}[1]{{  {\partial} \over {\partial {#1}}  }}
\newcommand{\papara}[1]{{ 
 {\stackrel{\raise.1ex\hbox{$\leftarrow$}}{\partial}   } 
\over {   \partial {#1}}  }}

\newcommand{\ddr}[1]{{ 
 {\stackrel{\raise.1ex\hbox{$\leftarrow$}}{\delta^r}   } 
\over {   \delta {#1}}  }}
\newcommand{\ddl}[1]{{ 
 {\stackrel{\lower.3ex \hbox{$\rightarrow$}}{\delta^l}   }
 \over {   \delta {#1}}  }}
\newcommand{\dd}[1]{{  {\delta} \over {\delta {#1}}  }}
\newcommand{\pa}{\partial}
\newcommand{\sokkel}[1]{\!  {\lower 1.5ex \hbox{${\scriptstyle {#1}}$}}}  
\newcommand{\larrow}[1]{\stackrel{\rightarrow}{#1}}
\newcommand{\rarrow}[1]{\stackrel{\leftarrow}{#1}}
\newcommand{\twobyone}[2]{\left(\begin{array}{c}{#1} \cr
                                   {#2} \end{array} \right)}
\newcommand{\twobytwo}[4]{\left[\begin{array}{ccc}{#1}&&{#2} \cr
                                  {#3} && {#4} \end{array} \right]}
\newcommand{\fourbyone}[4]{\left(\begin{array}{c}{#1} \cr{#2} \cr{#3} \cr
                                   {#4} \end{array} \right)}

\newcommand{\eq}[1]{{(\ref{#1})}}
\newcommand{\Eq}[1]{{eq.~(\ref{#1})}}
\newcommand{\Ref}[1]{{Ref.~\cite{#1}}}
\newcommand{\mb}[1]{{\mbox{${#1}$}}}
\newcommand{\equi}[1]{\stackrel{{#1}}{=}}
\newcommand{\succeqq}{\succeq}
\newcommand{\ccdot}{\cdot}
\newcommand{\qqa}{}
\newcommand{\qqb}{}
\newcommand{\qqc}{}
\newcommand{\qqd}{}
\newcommand{\yyy}{y}
\newcommand{\DD}{{\cal D}}

\newcommand{\proofbox}{\begin{flushright}
${\,\lower0.9pt\vbox{\hrule \hbox{\vrule
height 0.2 cm \hskip 0.2 cm \vrule height 0.2 cm}\hrule}\,}$
\end{flushright}}


\begin{titlepage}

\title{
\normalsize
\rightline{UFIFT-HEP-99-18}
\rightline{hep-th/9912017}
\vspace{2.5 cm}
\Large\bf Family of Boundary Poisson Brackets\\ 
}

\author{{\sc K.~Bering}
\footnote{Email address: {\tt bering@phys.ufl.edu, bering@nbi.dk}}
\\
Institute for Fundamental Theory\\ Department of Physics\\
University of Florida\\Florida 32611, USA\\
} 

\date{November 1999}

\maketitle
\begin{abstract}
We find a new $d$-parameter family of ultra-local boundary Poisson
brackets that satisfy the Jacobi identity. The two already known cases 
(hep-th/9305133, hep-th/9806249 and hep-th/9901112) of ultra-local 
boundary Poisson brackets are included in this new continuous family 
as special cases.
\end{abstract}

\bigskip

\vspace*{\fill}

\noi
PACS number(s): 02.70.Pt, 11.10.Ef.

\end{titlepage}
\vfill
\newpage

\setcounter{equation}{0}
\section{Introduction}

\vs
\noi
We have seen an increasing number of theories during the last few years
where boundaries or topological defects play a central role. Open 
strings ending on D-branes are one of the more recent examples and
surface terms in gravity is another. 

\vs
\noi
Typically a physical system has to fulfill extra constraints at a
boundary. There can be sound physical motives for imposing these
constraints (for instance local conservation of a quantity), but 
they can also appear for more ad hoc reasons.  

\vs
\noi
The work of \cite{soloviev,bering} use a generalized notion of 
functional differentiability, that led to two new boundary Poisson
brackets. They generalize the Poisson bracket of
Lewis, Marsden, Mongomery and Ratiu \cite{lmmr}.
It would be worthwhile to go back 
and re-examine various physical systems in this framework. 
It might lead to new ways of imposing (or {\em not} imposing!) 
boundary conditions and solving the system.

\vs
\noi
In this letter we shall stay in a general canonical formalism, 
and develop the Poisson brackets wrt.\ this extended notion of
functional differentiability.

\setcounter{equation}{0}
\section{Review of Boundary Poisson Brackets}

\vs
\noi
Consider a \mb{d+1} dimensional space-time \mb{\Sigma \times \R}, 
where space \mb{\Sigma} is a region of \mb{\R^d}, with a spatial
boundary \mb{\pa \Sigma}. Consider a phase space of (bosonic)
coordinate and momenta field variables \mb{\phi^{A}(x,t)},
\mb{A=1,\ldots,2N}. Time plays no role in the following, so we shall 
suppress $t$ in our formulae. We denote the non-degenerate symplectic
structure by \mb{\omega^{AB}}, which we for simplicity take to be 
ultra-local and constant.

\vs
\noi
Our building blocks for the boundary Poisson 
bracket\cite{soloviev,bering,solocomp} 
are the tower of higher Euler-Lagrange derivatives 
\beq
\dedetwo{F}{\phi^{A(k)}(x)}~,
\eeq
of a functional \mb{F}. 
(See for instance Olver \cite[p.365]{olver}.)
They have the property that
\beq
\delta F ~=~ \int_{\Sigma} d^{d}x ~ \sum_{k=0}^{\infty} \pa^{k} \left[ 
\dedetwo{F}{\phi^{A(k)}(x)} \delta \phi^{A}(x) \right]
\label{highfuncderiv}
\eeq
for arbitrary infinitesimal variations of the fields
\mb{\phi^{A}(x) \to \phi^{A}(x)+ \delta \phi^{A}(x)}.
The case \mb{k=0} corresponds to the usual Euler-Lagrange derivative.
Note, that the terms with \mb{k\neq 0}, by the divergence theorem, can 
be recast into an integral over the boundary \mb{\pa \Sigma}.
Clearly our ability to probe the higher derivatives diminishes as we 
constrain the dynamical fields \mb{\phi^{A}(x)} with more boundary 
conditions. Here we want to investigate the maximal effect of
the boundary terms, and hence we shall {\em not} impose any boundary 
conditions. (Needless to say that if boundary conditions at a later
stage become necessary, for instance during quantization, this will cause
no inconsistency, because it just restricts the number of field 
configurations.)

\vs
\noi
With this in mind, it is easy to see that the ``bulk'' Poisson bracket
\beq
  \{F , G \}_{(0)}~\equiv~\int_{\Sigma}d^{d}x 
\dedetwo{F}{\phi^{A(0)}(x)} \omega^{AB} \dedetwo{G}{\phi^{B(0)}(x)} 
\label{naivepb}
\eeq
built out of the usual Euler-Lagrange derivatives does not generically
satisfy the Jacobi identity: There can be a 
non-zero total derivative term left over. It is natural to ask if it
is possible to modify the ``bulk'' Poisson bracket \eq{naivepb} with 
a boundary term such that the Jacobi identity is restored identically.

\setcounter{equation}{0}
\section{A $d$-Parameter Family of Boundary Brackets}
\label{secnewpb}

\vs
\noi
We limit ourselves to the following ultra-local Ansatz for the full 
boundary Poisson bracket
\beq
 \{F , G \}~=~\sum_{k,\ell=0}^{\infty}c_{k\ell} \int_{\Sigma} d^{d}x ~
\pa^{k+\ell} \left[  \dedetwo{F}{\phi^{A(k)}(x)} 
\omega^{AB} \dedetwo{G}{\phi^{B(\ell)}(x)} \right]~,
\label{newpbansatz}
\eeq
where \mb{c_{k\ell}} is a sequence of constant coefficients. 
The ``bulk'' coefficient \mb{c_{00}\equiv 1} by definition. 
Soloviev \cite{soloviev} found that 
\beq 
\forall k,\ell:~c_{k\ell}~=~1
\label{solosolu}
\eeq
is a solution. Recently, we found another solution \cite{bering}
\beq
c_{k\ell}~=~\delta_{{\rm min}(k,\ell),0}
~=~\left\{\begin{array}{l} 1~~{\rm if}~~k=0~~ {\rm or} ~~\ell=0 \cr
      0~~ {\rm otherwise}~. \end{array} \right. 
\label{beringsolu}
\eeq
Our main new result is that 
\beq
c_{k\ell}(s)~=~\frac{(s)_{k}(s)_{\ell}}{(s)_{k+\ell}}~=~
\frac{\Gamma(k\!+\!s)\Gamma(\ell\!+\!s)}
{\Gamma(k\!+\!\ell\!+\!s)\Gamma(s)}~=~
\frac{B(k\!+\!s,\ell\!+\!s)}{B(k\!+\!\ell\!+\!s,s)}
\label{mainresult}
\eeq
is a solution for arbitrary complex parameter 
\mb{s\in \left((\C \cup \{\infty\}) \backslash (-\N)\right)^{d}} 
on $d$ copies of the Riemann sphere except for the negative integers
\mb{s \in (-\N)^d}, \mb{\N \equiv \{1,2,3,\ldots\}}, where some of the 
coefficients have poles. Here \mb{(s)_{n}=\Gamma(s+n)/\Gamma(s)} 
is the Pochhammer symbol in $d$ dimensions.

\vs
\noi 
The two previously found solutions \eq{solosolu} and \eq{beringsolu} 
correspond to \mb{s=\infty} and \mb{s=0}, respectively.

\setcounter{equation}{0}
\section{{} $x$-pointwise Poisson Bracket}

\vs
\noi
We have assumed that all relevant functionals are of the {\em local} form 
\beq
   F~=~\int_{\Sigma} d^{d}x ~f(x)~, 
\label{locfunc}
\eeq  
for some function \mb{f(x)~\equiv ~f\left(\pa^{k}\phi(x),x \right)},
that can depend on the dynamical fields \mb{\phi^{A}(x)} and on its spatial
derivative \mb{\pa^{k}\phi^{A}(x)} up to a finite order \mb{N}.

\vs
\noi
The notion of higher functional derivatives, if defined
merely from the descriptive property \eq{highfuncderiv}, is not unique. 
We emphasize that we use the canonical choice of the higher 
{\em Euler-Lagrange} derivatives:
\beq
\dedetwo{F}{\phi^{A(k)}(x)}~=~ 
E_{A(k)}f(x)~\equiv~ \sum_{m \geq k} \twobyone{m}{k} 
(-\pa)^{m-k} P_{A(m)}f(x)~,
\label{highel}
\eeq
where \mb{P_{A(m)}f(x)} denotes the partial derivative of \mb{f(x)} 
wrt.\ \mb{\pa^m\phi^{A}(x)}. It is easy to see that they obey 
property \eq{highfuncderiv}. The $x$-pointwise Poisson bracket reads
\beq
  \{f, g\}(x)~=~\sum_{k,\ell=0}^{\infty} c_{k\ell}~
\pa^{k+\ell} \left[ E_{A(k)}f(x) ~\omega^{AB}~ E_{B(\ell)}g(x) \right] ~.
\label{fullpbx}
\eeq

\setcounter{equation}{0}
\section{Fourier Transformed Bracket}

\vs
\noi
It is convenient to resum the higher derivatives in a series,
\bea
  P_{A}(q)f&\equiv&\sum_{k=0}^{\infty} q^{k}~P_{A(k)}f~, \cr
E_{A}(q)f& \equiv &\sum_{k=0}^{\infty} q^{k}~E_{A(k)}f~
~=~\exp\left[-\pa \papa{q}\right]P_{A}(q)f~,
\label{peq}
\eea
and to introduce the Fourier transform
\bea
P_{A}(y)f&\equiv&\int d^{d}q~e^{-qy} P_{A}(q)f~, \cr
E_{A}(y)f&\equiv&\int d^{d}q~e^{-qy} E_{A}(q)f 
~=~e^{-\pa y} P_{A}(y)f~.
\label{pey}
\eea
The Ansatz \eq{newpbansatz} for the boundary Poisson bracket becomes
of the form
\beq
  \{f, g\}~=~ \int d^{d}y~d^{d}y_{A}~d^{d}y_{B}~T(y,y_{A},y_{B})~e^{\pa
y}\left[ E_{A}(y_{A})f ~\omega^{AB}~ E_{B}(y_{B})g \right]
\label{fourierpb}
\eeq
for some kernel function \mb{T(y,y_{A},y_{B})}.
The $d$-parameter solution \eq{mainresult} can be written
\beq
 T(y,y_{A},y_{B})~=~  \int d^{d}q~e^{-qy}~
\Phi_{2}\left(s,s;s \left|q y_{A},q y_{B}\right.\right) 
\label{fouriermainresult}
\eeq
where \mb{\Phi_{2}} is a confluent hypergeometric function in two variables
(in $d$ dimensions):
\beq
\Phi_{2}\left(\mu,\nu;\lambda \left|x,y\right.\right) 
\equiv  \sum_{k,\ell=0}^{\infty}
\frac{(\mu)_{k}(\nu)_{\ell}}{(\lambda)_{k+\ell}}\frac{ x^{k}}{k!}
 \frac{ y^{\ell}}{\ell!}~, 
\eeq
where
\beq
\forall i=1, \ldots,d:~~
\lambda_{i} \notin (-\N_{0}) ~\vee~ \mu_{i}=\nu_{i}=\lambda_{i}=0~, 
\eeq
and where \mb{ \N_{0} \equiv \{0,1,2,3,\ldots\}}.

\setcounter{equation}{0}
\section{Sufficient Condition for the Jacobi Identity}

\vs
\noi
To show that a bracket is a Poisson Bracket, the non-trivial 
step is to prove the Jacobi identity.
It follows in straightforwardly, similar to the derivation given
in Appendix B of \cite{bering}, that 
\beq
\int  d^{d}\tilde{y}~ T(y\!+\!y_{B},y_{A},y_{B}\!+\!\tilde{y})
 ~T(\tilde{y}\!+\!y_{C},y_{C},y_{D})
~-~(A \leftrightarrow D,B \leftrightarrow C )~=~0
\label{suffcond}
\eeq 
is a sufficient condition for the Jacobi identity of \eq{fourierpb}.
In our case, \eq{fouriermainresult} the condition \eq{suffcond} holds,
because it is an identity for the \mb{\Phi_{2}} function. After a \mb{y\to q} 
Fourier transformation it reads:
\bea
\int d^{d}\tilde{y}~d^{d}\tilde{q}~
e^{-q y_{B}}~\Phi_{2}\left(s,s;s\left|qy_{A},
q(y_{B}\!+\!\tilde{y})\right.\right)~ 
e^{-\tilde{q}(\tilde{y}+y_{C})}~
\Phi_{2}\left(s,s;s\left|\tilde{q}y_{C},
\tilde{q}y_{D}\right.\right)\cr 
~-~(A \leftrightarrow D,B \leftrightarrow C )~=~0~,~~~~~~~~~ && 
s \notin (-\N)^{d}~.~~~~~~~~
\label{oursuffcond}
\eea
This identity is the special case \mb{s=t}, of a more general identity
\bea
\int d^{d}\tilde{y}~d^{d}\tilde{q}~
e^{-q y_{B}}~\Phi_{2}\left(s,t;2t\!-\!s\left|qy_{A},
q(y_{B}\!+\!\tilde{y})\right.\right)~ 
e^{-\tilde{q}(\tilde{y}+y_{C})}~
\Phi_{2}\left(s,s;t\left|\tilde{q}y_{C},
\tilde{q}y_{D}\right.\right) \cr
~-~(A \leftrightarrow D,B \leftrightarrow C )&=&0~,
\label{oursuffcondgen}
\eea
which is defined for pairs \mb{(s,t)} satisfying
\beq
\forall i=1, \ldots,d:~~
t_{i}, 2t_{i}\!-\!s_{i} \notin (-\N_{0}) ~~\vee~~ 
s_{i}=t_{i}=0~.
\label{oursuffdefregion}
\eeq

\vs
\noi
{\bf Acknowledgements}.
The research is supported by DoE grant no.~DE-FG02-97ER-41029.
We would like to thank J.~Rozowsky and  B.D.~Baker for carefully 
reading the manuscript.

\begin{appendix}

\setcounter{equation}{0}
\section{Proof of the \mb{\Phi_{2}} Identity \eq{oursuffcondgen}}

\vs
\noi
{}For completeness we provide a proof for 
the \mb{\Phi_{2}} identity \eq{oursuffcondgen}. Let us assume 
that we are given a pair \mb{(s,t)} satisfying \eq{oursuffdefregion}.
It is enough to give the proof for 
\beq
 t\!-\!s ~\notin~ (-\N_{0})^{d} ~,
\label{gencase}
\eeq 
because once this case is proven, the remaining case would then
follow from a continuity argument. 
Assuming\footnote{It is rather remarkable that the case $s=t$, 
which is the case \eq{oursuffcond} that we ultimately are interested in, 
is excluded by this assumption!} 
\eq{gencase}, we can rewrite the \mb{\Phi_{2}} functions as
\bea
\Phi_{2}\left(s,t;2t\!-\!s\left|qy_{A},
q(y_{B}\!+\!\tilde{y})\right.\right) 
\!\!\!\!&=&\!\!\!\!\int \!d^{d}\bar{y}_{A}~d^{d}\bar{q}_{A}~
e^{-\bar{q}_{A}\bar{y}_{A}}~
\Phi\left(s;t\!-\!s\left|\bar{q}_{A}y_{A}\right.\right)~
\beta\left(t\!-\!s,t\left|  q\bar{y}_{A},
q(y_{B}\!+\!\tilde{y})  \right.\right) ~, \cr \cr
\Phi_{2}\left(s,s;t\left|\tilde{q}y_{C},
\tilde{q}y_{D}\right.\right)
\!\!\!\!&=&\!\!\!\!\int \!d^{d}\bar{y}_{D}~d^{d}\bar{q}_{D}~
e^{-\bar{q}_{D}\bar{y}_{D}}~
\beta\left(s,t\!-\!s\left|\tilde{q}y_{C},
\tilde{q}\bar{y}_{D} \right.\right)~ 
\Phi\left(s;t\!-\!s\left|\bar{q}_{D}y_{D}\right.\right)~.
\label{phi2phibeta}
\eea

\vs
\noi
Here \mb{\Phi} is the usual confluent hypergeometric function 
in one variable (also known as \mb{{}_{1}F_{1}})
\beq
\Phi\left(\mu;\nu\left|x\right.\right)\equiv  \sum_{k=0}^{\infty}
\frac{(\mu)_{k}}{(\nu)_{k}}\frac{ x^{k}}{k!}~,~~~~~~~~~
\nu \notin (-\N_{0})^{d}~.
\eeq
We have introduced a convenient notation
\beq
\beta\left(\mu,\nu\left|x,y\right.\right) 
~\equiv ~\Phi_{2}\left(\mu,\nu;\mu\!+\!\nu\left|x,y\right.\right)
  ~=~\sum_{k,\ell=0}^{\infty}
\frac{B(\mu\!+\!k,\nu\!+\!\ell)}{B(\mu,\nu)}\frac{ x^{k}}{k!}
 \frac{ y^{\ell}}{\ell!}~,~~~~~~~~~
\mu\!+\!\nu \notin (-\N_{0})^{d}~,
\label{betadef}
\eeq
for a special case of the confluent hypergeometric function \mb{\Phi_{2}}.
We choose the name \mb{\beta} because of its relationship with the
Euler Beta function. Of crucial importance is the Kummer transformation
\beq
e^{z}\beta\left(\mu,\nu\left|x,y\right.\right) ~=~
\beta\left(\mu,\nu\left|x\!+\!z,y\!+\!z\right.\right)~, 
\eeq
which can easily be deduced from the integral representation:
\beq
\beta\left(\mu,\nu\left|x,y\right.\right) 
~=~\frac{1}{B(\mu,\nu)}\int_{0}^{1}
du~u^{\mu-1}(1-u)^{\nu-1}e^{xu+y(1-u)}~,~~~~~~~
{\rm Re}(\mu), {\rm Re}(\nu) > 0~. 
\label{betaint}
\eeq

\vs
\noi
We insert the expressions \eq{phi2phibeta} into equation 
\eq{oursuffcondgen}, then we apply a suitable Kummer transformation on
each of the two \mb{\beta} functions and finally we do a
translation of the integration variables
\beq
\bar{y}'_{A}~=~\bar{y}_{A}-y_{B}~,~~~~~
\bar{y}'_{D}~=~\bar{y}_{D}-y_{C}~.
\eeq
Then equation \eq{oursuffcondgen} becomes
\bea
\int d^{d}\tilde{y}~d^{d}\tilde{q}~
d^{d}\bar{y}'_{A}~d^{d}\bar{q}_{A}~
d^{d}\bar{y}'_{D}~d^{d}\bar{q}_{D}~
e^{-\bar{q}_{A}(\bar{y}'_{A}+y_{B})}~
\Phi\left(s;t\!-\!s\left|\bar{q}_{A}y_{A}\right.\right)~\cr\cr
e^{-\tilde{q}\tilde{y}}~
\beta\left(t\!-\!s,t\left|  q\bar{y}'_{A},
q\tilde{y}  \right.\right)~
\beta\left(s,t\!-\!s\left|0,\tilde{q}\bar{y}'_{D} \right.\right)~\cr\cr
e^{-\bar{q}_{D}(\bar{y}'_{D}+y_{C})}~
\Phi\left(s;t\!-\!s\left|\bar{q}_{D}y_{D}\right.\right)
~-~(A \leftrightarrow D,B \leftrightarrow C )&=&0~.
\eea
This is true, because
\beq
\int d^{d}\tilde{y}~d^{d}\tilde{q}~e^{-\tilde{q}\tilde{y}}~
\beta\left(t\!-\!s,t\left|  q\bar{y}'_{A},
q\tilde{y}  \right.\right)~
\beta\left(s,t\!-\!s\left|0,\tilde{q}\bar{y}'_{D} \right.\right)
~=~\Phi_{2}\left(t\!-\!s,t\!-\!s;2t\!-\!s
\left|q\bar{y}'_{A},q\bar{y}'_{D} \right.\right)~,
\eeq
so that the \mb{(A \leftrightarrow D,B \leftrightarrow C )} symmetry becomes
manifest.
\proofbox

\end{appendix}




\end{document}